\documentclass[aps,floats,epsf,twocolumn]{revtex4}
 \usepackage{graphics}

\begin{document}

\title{SO(3) symmetry between N\'{e}el and ferromagnetic order parameters for graphene in a magnetic field}
\author{Igor F. Herbut}

\affiliation{Department of Physics, Simon Fraser University,
 Burnaby, British Columbia, Canada V5A 1S6}

\begin{abstract}
I consider the Hubbard model of graphene in an external magnetic
field and in the Hartree-Fock approximation. In the continuum limit,
the ground state energy at half filling becomes nearly symmetric
under rotations of the three-component vector $\vec{\Delta}=(N_1,
N_2, m)$, with the first two components representing the N\'{e}el
order parameter orthogonal to and the third component the
magnetization parallel with the external magnetic field. When the
symmetry breaking effects arising from the lattice, Zeeman coupling,
and higher Landau levels are included the system develops a quantum
critical point at which the antiferromagnetic order disappears and
the magnetization has a kink. The observed incompressible states at
filling factors $\pm 1$ are argued to arise due to a finite third
component of the N\'{e}el order parameter at these electron
densities. Recent experiments appear consistent with $N_1=N_2=0$, and $N_3\neq 0$, at the filling factors
zero and one, respectively.
\end{abstract}
\maketitle

\vspace{10pt}

\section{Introduction}

The nature of the ground state of two-dimensional carbon (i. e.
``graphene") in a uniform magnetic field is an issue which has
attracted a lot of attention lately. The interest mainly stems from
the recent experimental observation \cite{zhang} of the additional
plateaus in the Hall conductivity at the values of zero and unity
which are not naturally explained by the picture of non-interacting
electrons alone \cite{ando}, \cite{sharapov}, \cite{guinea}. The
mechanism behind the formation of the gaps in the energy spectrum
that apparently develop within the zeroth Landau level has been a
matter of debate. On one hand, the Coulomb repulsion may on general
grounds be expected to favor breaking of the sublattice
\cite{gusynin}, or ``valley" symmetry \cite{nomura}, \cite{goerbig},
\cite{apalkov}, which in the zeroth Landau level are equivalent. If
the Zeeman coupling of the electron spin to the magnetic field is
entirely neglected the sense of this symmetry breaking can be equal
or opposite for the two projections of spin, depending on details of
the interaction on the lattice scale. This would yield a staggered
pattern of either charge or spin in the ground state, with the
concomitant many-body gap in the quasiparticle spectrum
\cite{herbut1}. One could, however, also imagine the opposite limit
of a strong coupling of the electron spin to the magnetic field, in
which the gap in the spectrum at half filling would become
essentially a single-particle Zeeman gap. Of course, such a
ferromagnetic ground state is already favored by the repulsive
Coulomb interaction alone, due to the familiar physics of Hund's rule \cite{alicea}. The interplay between
different possible instabilities in graphene in magnetic field
represents an important unsolved problem at the moment \cite{fuchs},
\cite{kun}. Here I would like to point out and explore a surprising symmetry
between different order parameters which may help its resolution.

\begin{figure}[t]\label{flowFig}
{\centering\resizebox*{65mm}{!}{\includegraphics{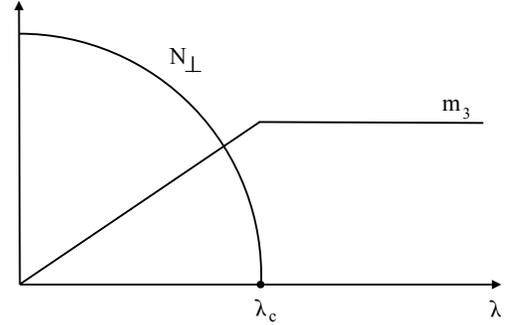}}
\par} \caption[] {The dependence of the N\'{e}el order parameter ($N_\perp $) orthogonal to  and the magnetization ($m_3$) parallel with the external magnetic field on the Zeeman energy ($\lambda$) at the filling factor zero (Eqs. (15)-(18)). The critical point is located at $\lambda_c= 2(g_a- g_f)/(\pi l^2)$, where $g_a$ and $g_f$ are the coupling constants in the N\'{e}el and the ferromagnetic channels, respectively. $l$ is the magnetic length.}
\end{figure}

  I consider the minimal Hubbard model for the interacting electrons living on a honeycomb lattice.
  Although admittedly a simplification, the Hubbard model already
contains most of the relevant
  physics. Without a magnetic field it exhibits both
  the  ferromagnetic and  antiferromagnetic ground states in its
  phase diagram \cite{peres}, with the latter as stable at half filling \cite{herbut2}.
  In a field, constraining the N\'{e}el vector to lie parallel to it leads to a discontinuous transition between the
  antiferromagnetic and ferromagnetic ground states at a critical Zeeman
  coupling, as discussed in the previous work by the author \cite{herbut1}.
  Here I calculate the ground state energy in the
  Hartree-Fock approximation, but for an {\it arbitrary} direction of the
  N\'{e}el vector. An immediate, and maybe not entirely unexpected result is
   that the ground state energy at half filling is minimized when the N\'{e}el order
  parameter $\vec{N}$ is in fact {\it orthogonal} to the external magnetic field.
  Restricting therefore $\vec{N}$ to its easy plane $(N_3 =0)$ and taking the magnetization
  $\vec{m}$ to be parallel to the external magnetic field $(m_1 = m_2 = 0)$,
  the ground state energy at half filling becomes nearly symmetric with respect
  to an internal $SO(3)$ group of rotations of the three-dimensional
  order parameter $\vec{\Delta}=(N_1, N_2, m_3)$. This symmetry
  becomes exact in the zeroth Landau level approximation, if the
  Zeeman effect and the discreteness of the lattice are also neglected. With these realistic symmetry breaking effects
  included one finds that with the increase of Zeeman coupling the system at half filling suffers a
  continuous quantum phase transition from the ``mixed"
  state with both $\vec{N}_\perp \neq 0$ and $m_3 \neq 0$ to a pure ferromagnet with
  $\vec{N}_\perp =0$ and $m_3 \neq 0$. This situation is depicted in  Fig. 1. With a change in the
  chemical potential the third N\'{e}el component $N_3$ eventually becomes finite and produces
 additional incompressible states at the filling factors $\pm 1$.
An observation of such, albeit according to our estimate, a rather
weak N\'{e}el order along the magnetic field at $\pm 1$ would
provide a direct support for our theory.

  Motivated by a recent experiment \cite{jiang}, I examine the dependence of the activation gap
  at filling $\pm 1$ on the in-plane component of the magnetic field, at the fixed perpendicular
  field. The present theory predicts such a gap to be completely independent of the in-plane component only if at the filling zero $N_\perp =0$ and $m_3\neq 0$, in which case it is simply given by $2 N_3$. Given that this seems to be the situation in the experiment, we conclude that for the experimentally relevant values of the parameters the ground state should have a finite $N_3$ at the filling factor $\pm 1$, whereas it
  should be a pure ferromagnet at zero filling. This conclusion is in accord with the observed dissipative nature of the state at the filling zero, which can be understood in terms  of the particular edge-state transport implied by  such a ferromagnetic ground state \cite{abanin}. It is then also in agreement with one of the possible scenarios discussed  previously in ref. (9), where the N\'{e}el order parameter was constrained to lie only along the magnetic field.

  The paper is organized as follows. In the next section I derive the expression for
  the ground-state energy of the system as a function of N\'{e}el and uniform magnetizations,
  in the Hartree-Fock approximation. In the section III the emergence and the breaking of the approximate SO(3) symmetry by different terms in the Hamiltonian is discussed. The discussion of the result as they pertain to experiment is given in the section V. Finally, the main results are summarized in section VI.

\section{Hartree-Fock ground-state energy}

Let me define the Hubbard model on a honeycomb lattice as $H_t
+H_U$, where
\begin{equation}
H_t= -t \sum_{\vec{A}, i, \sigma=\uparrow, \downarrow} u^\dagger
_\sigma (\vec{A}) v_\sigma (\vec{A}+\vec{b}_i) + H. c.,
\end{equation}
and
 \begin{eqnarray}
H_U= \frac{U}{16} \sum_{\vec{A}} [ (n(\vec{A} ) + n(\vec{A}+\vec{b})
)^2 + (n(\vec{A} ) - n(\vec{A}+\vec{b}) )^2 \\
\nonumber - (\vec{m}(\vec{A} ) +
 \vec{m}(\vec{A}+\vec{b}) )^2 - ( \vec{m}(\vec{A} ) -
\vec{m} (\vec{A}+\vec{b}) )^2 ].
\end{eqnarray}
The sites $\vec{A}$ denote one triangular sublattice of the
honeycomb lattice, generated by linear combinations of the basis
vectors $\vec{a}_1= (\sqrt{3}, -1)(a/2)$, $\vec{a}_2 = (0,a)$. The
second sublattice is then at $\vec{B}=\vec{A}+\vec{b}$, with
$\vec{b}= (1/\sqrt{3},1) (a/2)$. $a$ is the lattice spacing, and
$n(\vec{A}) = u^\dagger_\sigma (\vec{A})  u _{\sigma} (\vec{A}) $
and $\vec{m} (\vec{A})  = u^\dagger_\sigma (\vec{A})
\vec{\sigma}_{\sigma \sigma'}  u _{\sigma'} (\vec{A}) $ are the
particle number and the magnetization vector at site $\vec{A}$.
Variables at the second sublattice are analogously defined in terms
of fermion operators $v_\sigma (\vec{B}) $. It is easy to check that
$H_U$ is just the standard Hubbard on-site repulsion,  written here
in the rotationally invariant form which will prove to be suitable for our purposes.

  Here we are after the Hartree-Fock ground state of the Hubbard
  model in the uniform magnetic field and at half filling. Even at the highest laboratory fields
  $\sim 40T$ the magnetic length is much larger than the lattice spacing, $l\gg a$. It suffices therefore
  to consider only the continuum theory corresponding to the Hubbard model. General form of
  the low-energy field theory of graphene has been
  discussed in detail before \cite{herbut2}. For present purposes it will be enough to
  consider the Hartree-Fock ground state obtained after the
  usual decoupling of the interaction term in the ferromagnetic and
  antiferromagnetic channels. We thus assume a uniform density ($ \langle n
(\vec{X})\rangle = 1$, $\vec{X}=\vec{A}, \vec{B}$), and allow for
both  the uniform ($\vec{m} = \langle \vec{m}( \vec{A} ) + \vec{m}
(\vec{A}+\vec{b})\rangle$) and staggered magnetizations ($ \vec{N} =
\langle \vec{m}( \vec{A} ) - \vec{m} (\vec{A}+\vec{b})\rangle $).
Standard manipulations \cite{herbut3} give the ground state energy
per unit area and at half-filling to be
\begin{equation}
E = \frac{ \vec{N}^2}{4 g_a} + \frac{\vec{m}^2}{4 g_f} + E_0
[\vec{N}, \vec{m}],
\end{equation}
where $E_0$ is the ground state energy per unit area of the
resulting single-particle Hamiltonian
\begin{equation}
H_{HF}= I_2 \otimes H_0 - (\vec{N}  \cdot \vec{\sigma}) \otimes
\gamma_0 + ((\vec{\lambda} + \vec{m}) \cdot \vec{\sigma}) \otimes
I_4.
\end{equation}
$\vec{\lambda} = g_Z \vec{B}$ represents the Zeeman effect of the
uniform magnetic field $\vec{B}$, and $g_Z$ is the electron
g-factor. $H_0 = i \gamma_0 \gamma_i (-i
\partial _i - A_i)$, with $B=\epsilon_{3ij}\partial_i A_j$, is
the standard Dirac Hamiltonian near the two Fermi points in the
spectrum of $H_t$ \cite{semenoff}, \cite{carbotte}. For simplicity, the magnetic field will be
 assumed to be orthogonal to the graphene plane, until further notice. We work in
``graphene representation" of the Clifford algebra \cite{herbut2} in
which $\gamma_0 = I_2 \otimes \sigma_3$, $\gamma_1= \sigma_3 \otimes
\sigma_2$, $\gamma_2= -I_2 \otimes \sigma_1$, where $\{ I_2,
\vec{\sigma} \}$ is the standard Pauli basis of two-dimensional
matrices. Likewise, $I_4$ represents the four-dimensional unit
matrix. In our units $\hbar=e/c=v_F=1$. The coupling constants $g_f$
and $g_a$ are both proportional to the original repulsion energy
$U$, and positive. Although they are exactly equal at the lattice
scale, as evident from the form of $H_U$ in Eq. (2), they in general
will not be in the {\it effective} linearized Dirac theory sensible only
below a certain momentum cutoff $\Lambda$. Subscribing to the usual
logic of the low-energy description, $g_f$ and $g_a$ must be
considered only as effective parameters which themselves depend on
the cutoff $\Lambda$ in a way that ensures the cutoff independence
of all physical quantities. This observation will play a role in the
selection of the ground state, as will be discussed shortly.

  To proceed one needs the spectrum of the eight-dimensional
Hamiltonian $H_{HF}$. Perhaps this is most easily computed by
casting it into a block-diagonal form $H_{HF}= H_1 \oplus H_2$
\cite{herbut4}, with
 \begin{eqnarray}
  H_{1(2)} = \pm I_2 \otimes \sigma_1 (-i \partial_1 - A_1) - I_2 \otimes \sigma_2
  (-i \partial_2 - A_2)  \\ \nonumber
  - (\vec{N} \cdot \vec{\sigma})\otimes
  \sigma_3 + (\lambda + m_3 )  \sigma_3 \otimes I_2.
 \end{eqnarray}
Here we have chosen the third spin axis and the magnetization to be
along the magnetic field, i. e. $m_1=m_2=0$, $m_3 >0$. Next we note
that $H_1 = U_1 ^\dagger H U_1 $, with $U_1 = I_2 \oplus i\sigma_2$,
and
\begin{equation}
H= H_0 + i N_1 \gamma_0 \gamma_3 + i N_2 \gamma_0 \gamma_5 - N_3
\gamma_0 + i (\lambda + m_3) \gamma_3 \gamma_5
\end{equation}
and $\gamma_3= \sigma_1 \otimes \sigma_2$, $\gamma_5= \sigma_2
\otimes \sigma_2$. Similarly, $H_2 = U_2 ^\dagger H U_2 $, with $U_2
= i\sigma_2 \oplus I_2$, with $N_3\rightarrow -N_3$. It is
sufficient therefore to find the spectrum of the single
four-dimensional Hamiltonian $H$. After some algebra the eigenvalues
of $H_{HF}$ are this way found at $\pm e_{n\sigma}$ with
\begin{equation}
e_{n\sigma} = \{ N_\perp ^2 + [ (N_3 ^2 + 2 n B ) ^{1/2} + \sigma
(\lambda + m_3 ) ]^2 \}^{1/2},
\end{equation}
with $\sigma=\pm 1$, $\vec{N}_\perp=(N_1,N_2)$, and the degeneracies
per unit area of $ 1/(\pi l^2) $ and $1/(2\pi l^2) $, for
$n=1,2,3,...$ and $n=0$, respectively. Note that for $\vec{N}=0$ the
eigenvalues become the familiar relativistic Landau levels split by
the Zeeman term, as expected. For $\lambda + m_3 = 0$, on the other
hand, the spectrum reduces to the Landau levels of the massive Dirac
Hamiltonian, with the mass $|\vec{N}|$, also in agreement with the
previous calculations \cite{herbut1}.

  At half filling then the ground state of $H_{HF}$ has all the
  eigenstates with negative energies filled, and all the others
  empty, so
  \begin{equation}
  E_0 [\vec{N},m_3]= -\frac{1}{2\pi l^2} \sum_{\sigma=\pm 1} [
  e_{0 \sigma}
  +  2  \sum_{n\neq 0}  e_{n\sigma}  ] .
\end{equation}
The variational Hartree-Fock ground-state energy is then determined by the least value of
the expression in Eq. (3).

\section{SO(3) symmetry and its breaking}

Let us minimize the Hartree-Fock ground state energy given by Eqs.
(3) and (8), assuming $\lambda\neq 0$. Choosing $|\vec{N}|$ and
$N_3$ as independent variables and then differentiating with respect
to $N_3$ yields
\begin{equation}
\sum_{\sigma=\pm 1} \sigma (\lambda+m_3 )[ \frac{ 1} { e_{0\sigma}
} +
 \sum_{n\neq 0}\frac{2}{ e_{n\sigma}} \frac{  N_3  }{  \sqrt{N_3 ^2 + 2 n B}  }   ]=0.
  \end{equation}
  The left-hand side of the equation vanishes for $N_3=0$, and
  otherwise is a negative definite function of $N_3$. $N_3 =0$ is therefore the only
  solution. One can also show that this solution represents the {\it minimum} of the energy.
  Restricting the N\'{e}el vector then to be orthogonal to the
  magnetic field the ground state energy in Eq. (3) can be rewritten as,
  \begin{equation}
  E= E_{SO(3)} + E',
  \end{equation}
  where
  \begin{equation}
  E_{SO(3)} = \frac{\vec{N}_\perp ^2 + m_3 ^2}{4 g_a} - \frac{1}{\pi l^2} (\vec{N}_\perp ^2 + m_3 ^2 )^{1/2},
  \end{equation}
  and $E'= E_{\Lambda} + E_Z + E_{HLL}$, with
  \begin{equation}
E_{\Lambda} =  \frac{(g_a - g_f) }{4 g_f g_a}  m_3 ^2 ,
\end{equation}
\begin{equation}
E_Z = \frac{1}{\pi l^2}[ (\vec{N}_\perp ^2 + m_3 ^2 )^{1/2} - (
\vec{N}_\perp ^2 + (\lambda+ m_3 ) ^2 )^{1/2}] ,
\end{equation}
\begin{equation}
E_{HLL} = -\sum_{\sigma=\pm 1} \sum_{n=1} ^M \frac{ \{ \vec{N}_\perp
^2 + [(2 n B )^{1/2}  + \sigma( \lambda + m_3 ) ] ^2 \} ^{1/2} }{\pi
l^2}.
\end{equation}
This form makes it manifest that the Hartree-Fock ground state
energy at half filling is nearly symmetric with respect to rotations
of the three-component order parameter $\vec{\Delta}=(N_1, N_2,
m_3)$. The identification of this approximate internal $SO(3)$
symmetry is our central result. The $SO(3)$ symmetry is in the
Hubbard model broken by three terms of different physical origin: 1)
the difference between the coupling constants $g_a$ and $g_m$ at the
cutoff $\Lambda$ ($E_\Lambda$), 2) the finite Zeeman coupling to the
magnetic field  $\lambda$ ($E_Z$), and 3) the $n\neq 0$ Landau level
contribution ($E_{HLL}$). If one were to consider only the zeroth
Landau level, use the bare values of the couplings for which $g_a =
g_m$, and neglect the Zeeman coupling, the Hartree-Fock ground-state
energy would become perfectly $SO(3)$ symmetric. In particular, the
antiferromagnetic state with the N\'{e}el vector orthogonal to the
external magnetic field and the ferromagnetic state with the
magnetization along the same field would  in this approximation
appear as degenerate.

  Clearly, the Zeeman term favors magnetization.
  Similarly, it can be shown that $H_{HLL}$ prefers the N\'{e}el
components \cite{remark}.
  It is less obvious what the sign of the ``easy-axis anisotropy" , i. e. of $g_a-g_f$, in $E_\Lambda$ should be.
  This is determined by the flow of the two coupling constants as
  the high-energy modes between the momenta $\sim 1/l$  and $\sim 1/a$
  are integrated out essentially at zero magnetic field. Since the leading
  instability of the Hubbard model on a honeycomb lattice and at half filling as the
  interaction strength is increased is towards the antiferromagnetism,
  we may assume that in general $g_a > g_f$ in the low energy
  theory. This would also be in accord with the explicit renormalization group
  calculation for weak couplings \cite{herbut2}.

  The ground state in the Hartree-Fock approximation is thus the
    result of the competition between the
    high-energy modes represented by $H_\Lambda$ and $H_{HLL}$,
   and the Zeeman coupling, which favor antiferromagnetic and
   ferromagnetic solutions, respectively. Since the effect of
    $H_{HLL}$ is in the same direction as of $H_\Lambda$, to
    keep the algebra simple we will drop $H_{HLL}$ altogether and just
    assume $g_a-g_f >0$ in the Eq. (10). This is a particularly good approximation at the laboratory magnetic fields
     at which the inclusion of the higher Landau levels is expected to provide corrections to our results
     of higher order in the small parameter $a/l$.  Minimizing the energy in Eq. (10), for $\lambda <\lambda_c$
      we then find
    \begin{equation}
    |\vec{N}_\perp | = \frac{g_a }{g_a - g_f} \sqrt{\lambda_c ^2 - \lambda^2},
    \end{equation}
    \begin{equation}
    m= \frac{g_f}{g_a - g_f}  \lambda ,
    \end{equation}
    and for $\lambda > \lambda_c$,
    \begin{equation}
    \vec{N}_\perp =0,
    \end{equation}
    \begin{equation}
     m= \frac{g_f}{g_a - g_f}  \lambda_c ,
    \end{equation}

\noindent
    where $\lambda_c= 2(g_a - g_f)/ (\pi l^2)$. With the increase of
    the Zeeman coupling there is a continuous
    transition at which the N\'{e}el order disappears, and the
    magnetization saturates to its maximum (Fig. 1).

    Note that if the N\'{e}el vector is constrained to be along the magnetic field,
    so that $\vec{N}_\perp=0$, the above quantum critical point becomes replaced with
    the level crossing between the purely antiferromagnetic
    ($N_3 \neq 0$, $m_3 =0$) and purely ferromagnetic ($N_3=0$, $m_3\neq
    0$) ground states \cite{herbut1}. Such a pure antiferromagnet would correspond
    to a local maximum of the energy in Eq. (3), however. It seems always
    energetically favorable to compromise between the two competing magnetic orderings by
    orienting the N\'{e}el vector orthogonally to the field.

\section{Discussion}

    While it is not {\it a priori} clear at which side of the transition
    should the experimental samples lie at half filling, increasing sufficiently the
    component of the magnetic field parallel to the graphene layer
    should always place the system into the purely ferromagnetic
    state. The experimental fact that the system at the filling factor zero is dissipative suggests that
    this is probably already the case even for the magnetic field completely orthogonal to the graphene's plane
    \cite{abanin}. Spin splitting as the  origin of the measured gap at half filling
    has also been suggested very recently in ref. \cite{jiang}. Most importantly, the experiment of Jiang et al.
    offers compelling evidence that the gap at filling factor $\pm 1$ is
    independent of the total magnetic field, and thus likely to be a consequence of
    interactions \cite{jiang}. And this is precisely what
    follows from our theory \cite{herbut1}. When the chemical potential
    gets close to the energy of the first excited states laying at
    \begin{equation}
    \mu_c =  \pm \sqrt{ \vec{N} _\perp ^2 + (\lambda+m_3)^2 },
\end{equation}
 the Hartree-Fock ground state energy may always be lowered by developing a finite third N\'{e}el component
 $N_3$. This pushes half of the states in question below the chemical potential and opens a gap. This way an incompressible state at
filling factors $\pm 1$ would be formed irrespectively of whether
$\vec{N}_\perp$ was zero or finite at the filling factor zero. Note,
however, that the Eq. (7) implies that this mechanism is operative only in the $n=0$ Landau
level; the states belonging to the other Landau levels do not get
split but only shifted in energy when $N_3 \neq 0$. Consequently, at
weak coupling $N_3$ will become finite only in the $\pm 1$ state and
otherwise not. The incompressible states that follow from the
inclusion of a finite chemical potential into Eq. (3) lie therefore
only at filling factors $0$, $\pm 1$, and all even integers.

Let us assume therefore a week third N\'{e}el component $N_3$ at the filling factor
$\pm 1$. The activation gap at this filling then becomes
\begin{equation}
E_{gap}= \frac{2 (\lambda+m_3)}{( N_\perp ^2 + (\lambda+m_3)^2 )^{1/2}} N_3 + O(N_3 ^2).
\end{equation}
For a small Zeeman term, $\lambda<\lambda_c$, Eqs. (15) and (16) give then
\begin{equation}
E_{gap} = \frac{2\lambda}{\lambda_c} N_3 + O(N_3 ^2),
\end{equation}
whereas when $\lambda >\lambda_c$ and $N_\perp =0$,
\begin{equation}
E_{gap} = 2 N_3 + O(N_3 ^2).
\end{equation}
Recalling that both $N_3 \sim B_\perp$ and $\lambda_c \sim B_\perp$, whereas $\lambda\sim B$,
where $B$ is the total and $B_\perp$ the perpendicular component of the magnetic field, we see that the gap
is independent of the field's in-plane component {\it only} in the latter case. The experiment is
therefore consistent with $N_3\neq 0$ at the filling factor $\pm 1$, and with $N_\perp =0$ at the filling factor zero.
This may not be very surprising in view of the equality between the couplings $g_a$ and $g_f$ at the lattice scale, which makes the critical Zeeman coupling $\lambda_c$ likely to be exceeded by the experimental value of $\lambda$ \cite{comment}.

   Finally, restoring a finite range to electron-electron interactions may
   replace the antiferromagnetic ground state discussed here with a
   charge density wave with the particle density alternating (around the value of unity) on the
   two sublattices. In this case the mean field phase
   diagram would remain the same as discussed previously in ref. (9).
Provided that the on-site repulsion is indeed the dominant effect of
the Coulomb interaction in graphene, our prediction is that at
filing factors $\pm 1$ and at magnetic fields $B\sim 10T$ the system
has a weak N\'{e}el component $N_3 \sim (a/l)^2 \mu_B\sim (10^{-3} -
10^{-4}) \mu_B $ per electron. $\mu_B$ is the Bohr magneton. The
inherent weakness of the N\'{e}el order derives from a small
fraction of electrons occupying the relevant $n=0$ Landau level,
which is the sole source of antiferromagnetism in this case. The existence of such ordering at the fillings
$\pm 1$ can be used to distinguish the present proposal from all others in the current
 literature on the subject.

\section{Summary}

To summarize, the ground-state energy in the Hartree-Fock
approximation to the Hubbard model on honeycomb lattice and at half
filling is nearly symmetric with respect to rotations between the
components of the N\'{e}el order orthogonal to and the magnetization
parallel with the external magnetic field. The effects of the
symmetry breaking terms originating from discreetness of the
lattice, Zeeman interaction, and $n\neq 0$ Landau levels were
examined. It was shown that whereas the component of the N\'{e}el
order parameter parallel to the magnetic field always vanishes at
half filling, it becomes finite (only) at filling factors $\pm 1$,
introducing this way a gap in the spectrum. The existence of such a
weak N\'{e}el order along the magnetic field $\sim 10^{-4} \mu_B$
per electron is proposed as a litmus test of the present theory.

\section{Acknowledgments}

This work was supported by NSERC of Canada. The author is grateful
to Asle Sudbo and the Center for Advanced Study at the Norwegian
Academy of Science and Letters in Oslo for their hospitality during
the final period of completion of this work, and to Allan MacDonald
for useful discussions.


\begin{thebibliography}{99}
\bibitem{remark} More precisely, the minimum of $E_{HLL}$ alone for a fixed $|\vec{\Delta}|$ is at $m_3 \sim
|\vec{\Delta}| g_z g_a (a/l)^2\ll |\vec{N}| $.
\bibitem{comment} Note that when $N_\perp =0$ the activation gap at zero filling is $2 (\lambda+m_3)$, and thus, as demanded  by its observed magnitude, significantly enhanced by the interactions.
\end{thebibliography}
\end{document}